\documentclass[fleqn,usenatbib]{mnras}

\usepackage{newtxtext,newtxmath}
\usepackage[T1]{fontenc}

\DeclareRobustCommand{\VAN}[3]{#2}
\let\VANthebibliography\thebibliography
\def\thebibliography{\DeclareRobustCommand{\VAN}[3]{##3}\VANthebibliography}

\usepackage{graphicx}	% Including figure files
\usepackage{amsmath}	% Advanced maths commands
\usepackage{bm}
\usepackage{hyperref}

\defcitealias{KT18}{KT18}

\newcommand{\aref}[1]{\hyperref[#1]{Appendix \ref{#1}}}

\usepackage{scalerel,tikz}
\usetikzlibrary{svg.path}
\definecolor{orcidlogocol}{HTML}{A6CE39}
\tikzset{orcidlogo/.pic={
\fill[orcidlogocol] svg{M256,128c0,70.7-57.3,128-128,128C57.3,256,0,198.7,0,128C0,57.3,57.3,0,128,0C198.7,0,256,57.3,256,128z};
\fill[white] svg{M86.3,186.2H70.9V79.1h15.4v48.4V186.2z}
svg{M108.9,79.1h41.6c39.6,0,57,28.3,57,53.6c0,27.5-21.5,53.6-56.8,53.6h-41.8V79.1z M124.3,172.4h24.5c34.9,0,42.9-26.5,42.9-39.7c0-21.5-13.7-39.7-43.7-39.7h-23.7V172.4z}
svg{M88.7,56.8c0,5.5-4.5,10.1-10.1,10.1c-5.6,0-10.1-4.6-10.1-10.1c0-5.6,4.5-10.1,10.1-10.1C84.2,46.7,88.7,51.3,88.7,56.8z};
}}
\newcommand\orcidicon[1]{\href{https://orcid.org/#1}{\mbox{\scalerel*{
\begin{tikzpicture}[yscale=-1,transform shape]
\pic{orcidlogo};
\end{tikzpicture}
}{|}}}}

\title[Metallicity correlations in Auriga simulations]{Cosmological evolution of metallicity correlation functions from the Auriga simulations}

\author[Z. Li et al.]{
Zefeng Li$^{\orcidicon{0000-0001-7373-3115}}$,$^{1,2,3}$\thanks{E-mail: zefeng.li@anu.edu.au}
Robert, J. J. Grand$^{\orcidicon{0000-0001-9667-1340}}$,$^{4}$
Emily Wisnioski$^{\orcidicon{0000-0003-1657-7878}}$,$^{1, 2}$
J. Trevor Mendel$^{\orcidicon{0000-0002-6327-9147}}$,$^{1, 2}$
Mark R. Krumholz$^{\orcidicon{0000-0003-3893-854X}}$,$^{1, 2}$
\newauthor Yuan-Sen Ting$^{\orcidicon{0000-0001-5082-9536}}$,$^{1, 2, 5}$
Ruediger Pakmor$^{\orcidicon{0000-0003-3308-2420}}$,$^{6}$
Facundo A. G\'omez$^{\orcidicon{0000-0003-4232-8584}}$,$^{7,8}$
Federico Marinacci$^{\orcidicon{0000-0003-3816-7028}\,9,10}$
and \newauthor Ioana Ciuc\u{a}$^{\orcidicon{0000-0001-6823-5453}\,1, 2, 5}$
\\
$^1$Research School of Astronomy \& Astrophysics, Australian National University, Weston Creek, ACT 2611, Australia\\
$^2$ARC Centre of Excellence for All Sky Astrophysics in 3 Dimensions (ASTRO 3D), Canberra, ACT 2611, Australia\\
$^3$Centre for Extragalactic Astronomy, Department of Physics, Durham University, South Road, Durham DH1 3LE, UK\\
$^4$Astrophysics Research Institute, Liverpool John Moores University, 146 Brownlow Hill, Liverpool L3 5RF, UK\\
$^5$Research School of Computing, Australian National University, Acton, ACT 2612, Australia\\
$^6$Max-Planck-Institut f\"ur Astrophysik, Karl-Schwarzschild-Str 1, D-85748 Garching, Germany\\
$^7$Departamento de Astronom\'ia, Universidad de La Serena, Av. Juan Cisternas 1200 Norte, La Serena, Chile\\
$^8$Instituto Multidisciplinario de Investigaci\'on y Postgrado, Universidad de La Serena, Av. Ra\'ul Bitr\'an 1305, La Serena, Chile\\
$^{9}$Department of Physics \& Astronomy `Augusto Righi', University of Bologna, Via Gobetti 93/2, I-40129 Bologna, Italy\\
$^{10}$INAF, Osservatorio di Astrofisica e Scienza dello Spazio Bologna, Via Gobetti 93/3, I-40129 Bologna, Italy\\
}

\date{Accepted XXX. Received YYY; in original form ZZZ}

\pubyear{2024}

\begin{document}
\label{firstpage}
\pagerange{\pageref{firstpage}--\pageref{lastpage}}
\maketitle

\begin{abstract}
We study the cosmological evolution of the two-point correlation functions of galactic gas-phase metal distributions using the 28 simulated galaxies from the Auriga Project. Using mock observations of the $z = 0$ snapshots to mimic our past work, we show that the correlation functions of the simulated mock observations are well matched to the correlation functions measured from local galaxy surveys. This comparison suggests that the simulations capture the processes important for determining metal correlation lengths, the key parameter in metallicity correlation functions. We investigate the evolution of metallicity correlations over cosmic time using the true simulation data, showing that individual galaxies undergo no significant systematic evolution in their metal correlation functions from $z\sim3$ to today. In addition, the fluctuations in metal correlation length are correlated with but lag ahead fluctuations in star formation rate. This suggests that re-arrangement of metals within galaxies occurs at a higher cadence than star formation activity, and is more sensitive to the changes of environment, such as galaxy mergers, gas inflows / outflows, and fly-bys.
\end{abstract}


%%%%%%%%%%%%%%%%%%%%%%%%%%%%%%%%%%%%%%%%%%%%%%%%%%

%%%%%%%%%%%%%%%%% BODY OF PAPER %%%%%%%%%%%%%%%%%%

% (It is noted in the same section) morph has already been explored. l_corr vs. mass / sfr remains invariant, and l_corr vs. cosmic time remains "invariant". 

\section{Introduction}

Understanding the baryonic processes responsible for metal (elements heavier than hydrogen and helium) transportation is crucial in galaxy evolution. Metals can be found in stars, their birth places where they are synthesized through nucleosynthesis; they can also be found in interstellar medium (ISM), into which they are injected by stars through various mechanisms including supernova explosions and stellar winds. Once in the ISM, metals mix with the surrounding gas and diffuse, leading to changes in the overall metallicity distribution within galaxies, participating in next-generation star formation, and thus forming a crucial part of the baryon cycle \citep{Tinsley80}.

Both observers and theorists have studied this process. On the observational side, gas-phase metallicity is most commonly measured by the abundance of oxygen, the most abundant metal in the Universe. The invention of integrated field units \citep[IFUs, e.g.][]{Croom12, Sanchez12, Bundy15, Erroz-Ferrer19} has made it possible to measure the spatial distribution of oxygen abundance in nearby galaxies \citep[e.g.][]{Rosales-Ortega11, Sanchez-Menguiano16_azi}. IFU studies have revealed that metallicity gradients are ubiquitous and that their steepness is correlated with other galaxy properties such as stellar mass \citep[e.g.][]{Belfiore17, Ho18, Poetrodjojo18, Sanchez-Menguiano18, Kreckel19}. On the theoretical side, numerical models have begun to explore the origin and evolution of metallicity gradients \citep[e.g.][]{DiMatteo09, MaX17, Sharda21, Tissera22}, but have also gone beyond simply 1D statistics such as gradients to examine chemical mixing in a broader view. The models contain various mechanisms, including bar driven mixing \citep{DiMatteo13}, spiral-arm driven mixing \citep{Grand16, Orr23}, thermal instabilities \citep{Yang12}, gravitational instabilities \citep{Petit15}, cosmological accretion \citep{Ceverino16}, and supernova-driven turbulence \citep{deAvillez02, Colbrook17}.

Exploring beyond simple metallicity gradients requires new statistical tools to characterise 2D metallicity maps. One of the simplest and most straightforward to measure from observations is the two-point correlation function. Two-point correlations are able to provide unique information of ISM turbulence and chemical mixing by decoding metal fields of galaxies. \citet[hereafter \citetalias{KT18}]{KT18} propose a model to predict this quantity that considers metal injection and diffusion. This prediction has inspired a number of observational studies \citep{Kreckel20, Metha21, Williams22} that measure the two-point correlations (or closely-related statistics) of PHANGS-MUSE galaxies. While much of this effort has focused on very nearby galaxies observed at extremely high resolution, more recently, \cite{Li21, Li23} extended the method to larger but more distant galaxy samples in CALIFA \citep{Sanchez12} and AMUSING++ \citep{Lopez-Coba20}, respectively, and demonstrated its robustness against beam smearing, choice of metallicity diagnostic, and binning used to remove the overall metallicity gradient. The samples in these studies are large enough to permit for the first time an examination of statistical distributions of metallicity correlation functions across the local galaxy population, and the correlations between metal distributions and other galactic properties. Most strikingly, these studies have uncovered a robust correlation between the galaxies' metallicity correlation lengths -- the characteristic size scales of their metallicity correlations after removing the large-scale gradient -- and their other bulk properties such as stellar mass, size, and star formation rate. This discovery raises immediate questions: how and when in the course of galaxy evolution were these relationships established? Have they varied over cosmic time?

While there has been extensive work on the cosmological evolution of metallicity gradients using cosmological zoom-in simulations \citep[e.g.][]{DiMatteo09, Torrey12, MaX17, Bellardini21, Metha23}, there has yet to be a similar study of higher-order metallicity statistics such as the metallicity correlation function, despite the growing body of observational literature. In this paper we present a pioneering study aimed at filling this gap. The correlation function is of interest because it encodes the fundamental physics of chemical mixing in the ISM, the process responsible for distributing metals from their birth sites. This process in turn depends on the evolution of ISM turbulence across cosmic time, another process of great interest. Our specific aims are to (1) determine if the simulations are able to reproduce the distribution of two-point correlation functions found in observed local galaxies, (2) use the simulations to study the cosmological evolution of two-point correlations so that we can understand when and how they are established, and (3) guide future observational work aimed at measuring metallicity correlations beyond the local Universe. 

This paper is organized as follows. In \autoref{sec:sim}, we describe the Auriga simulations and the procedures by which we extract metallicity maps from them. In \autoref{sec:mock}, we discuss the pipeline of mock observations we use to test whether the simulations are consistent with local galaxy observations. In \autoref{sec:res}, we describe our findings regarding the cosmological evolution and origin of metallicity correlations. Finally, we discuss and summarise our work in \autoref{sec:dis_sum}.

\section{Simulations}
\label{sec:sim}

\subsection{Description}
\label{sec:desc}

In this work, we examine 28 simulated galaxies from the Auriga Project \citep[][named as AuN, where N is the halo number]{Grand17}. The Auriga simulations are high-resolution cosmological zoom-in simulations using the magnetohydrodynamic (MHD) code \textsc{arepo} \citep{Springel10}. \textsc{arepo} is a quasi-Lagrangian, moving-mesh code that tracks the evolution of MHD cells and collisionless particles in a $\Lambda$CDM cosmological setting, with \citet{Planck14} cosmological parameters $\Omega_{\rm m}=0.307$, $\Omega_{\rm b}=0.048$, $\Omega_{\rm \Lambda}=0.693$, and a Hubble constant of $H_0 = 100h$ km s$^{-1}$ Mpc$^{-1}$, where $h=0.6777$. The simulations include primordial and metal-line cooling, a uniform ultraviolet (UV) background field for reionisation, star formation, magnetic fields, active galactic nuclei, energetic and chemical feedback from Type II supernovae, and mass loss/metal return due to Type Ia supernovae and asymptotic giant branch (AGB) stars, accounting for 9 elements \citep[H, He, C, N, O, Ne, Mg, Si, Fe;][]{Grand17}.

Star formation in the Auriga simulations occurs in cells that exceed a density threshold of $\rho_0 = 0.13$ atoms cm$^{-3}$; cells exceeding this threshhold form stars with a star formation timescale of $\tau_{\rm SF} = 2.2$ Gyr \citep[following][]{Springel03}. The Auriga simulations assume that every star cell represents a simple stellar population (SSP) with a specified age and metallicity that is directly inherited from the surrounding gas. The distribution of stellar masses present in each SSP follows a \cite{Chabrier03} initial mass function.

The Auriga simulation suite includes runs from four different particle resolutions, and in this work we discuss two of them. Six halos (Au6, Au16, Au21, Au23, Au24, and Au27) have been simulated with a baryonic mass resolution of $\sim6\times10^3$M$_{\odot}$, corresponding to maximum gravitational softening length of 184 physical pc; these are resolution level 3 in the Auriga terminology. By contrast 28 have a baryonic mass resolution of $\sim5\times10^4$M$_{\odot}$, corresponding to a maximum softening length of 369 physical pc (level 4); note that the 6 halos simulated at level 3 were also simulated at level 4. We perform a comparison of the results obtained from the six halos that are simulated at both resolutions in \aref{app:levels}. There we show that the results derived at the two resolutions are statistically similar, and thus for the bulk of our analysis in this paper we will use the 28 halos available at lower resolution due to the greater statistical power they offer. The exception is in \autoref{sec:mock}, where we construct mock observations, and the higher spatial resolution is useful for studying observational effects (e.g. beam smearing).

\begin{figure*}
\includegraphics[width=1.0\linewidth]{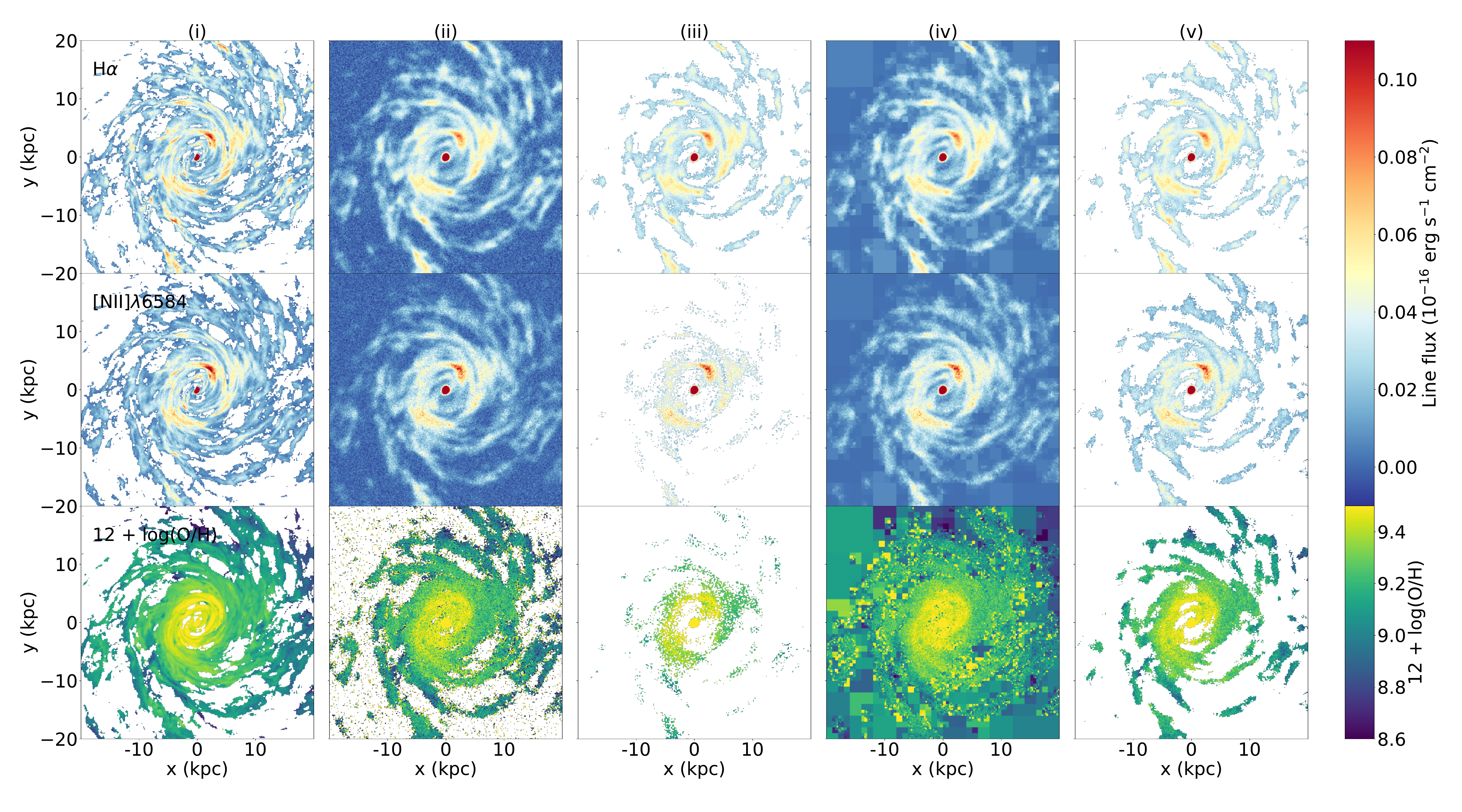}
\caption{Illustration of the mock observations pipeline applied to the galaxy Au6. Each column shows one step in the procedure described in \autoref{sec:mock_vs_obs}, respectively; these steps are (i) producing true emission line maps, (ii) convolution of the maps with a synthetic beam and addition of noise, (iii) pruning of low signal-to-noise pixels, (iv) reconstruction of the low signal-to-noise areas using the \textsc{adabin} algorithm; (v) masking the remaining low signal-to-noise regions of the adaptively-binned maps. From top to bottom the rows show H$\alpha$ maps, [N\textsc{ii}]$\lambda6584$ maps, and metallicity maps. The first two rows correspond to the colour bar on the upper right, showing line flux emitted per unit area, and the bottom row corresponds to the lower right colour bar, showing metallicity.}
\label{fig:mock_obs}
\end{figure*}

\begin{figure}
\includegraphics[width=1.0\linewidth]{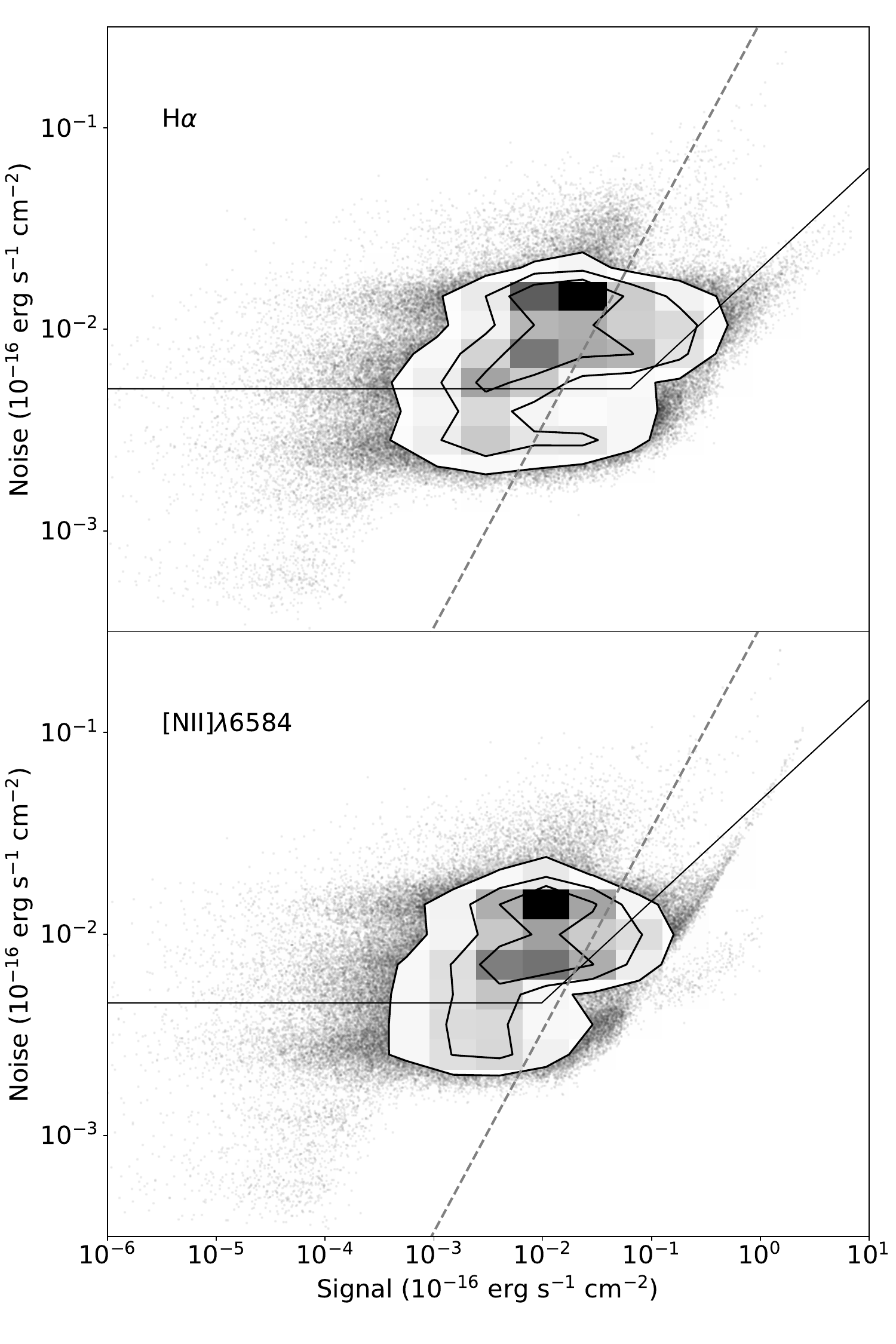}
\caption{Distribution of signal versus noise for the H$\alpha$ (upper panel) and [N\textsc{ii}]$\lambda6584$ (lower panel) lines over all pixels extracted from four sample galaxies in AMUSING++; see main text for details. The contours represent the areas that enclose 39\% ($1\sigma$ in 2D Gaussian), 68\% ($2\sigma$), and 86\% ($3\sigma$) of the data; outside the outermost contour, we show individual pixels as black points. The black solid lines show our best flat-plus-linear fit, while the gray dashed lines denote a fixed signal-to-noise ratio of 3.}
\label{fig:signal_noise}
\end{figure}

\subsection{Data extraction and analysis}
\label{sec:data3}

For each simulation snapshot we extract a box centred on the galactic centre for analysis; our boxes extend to $\pm 10$ kpc on either size of the galactic plane\footnote{In the Auriga simulations the $z$ axis, which by convention is normal to the galactic plane, is defined to be parallel to the total angular momentum vector of the star particles within $0.1R_{200}$ of the galactic centre, where $R_{200}$ is the virial radius. We have verified that alternative choices of definition of the $z$ axis, for example defining it based on dense gas cells rather than star particles, produces negligible changes in the results at both local and high-redshift Universe.}, but the choice of size in the directions parallel to the galactic plane requires some care. For the purposes of our mock observations of $z=0$ galaxies (\autoref{sec:mock}) we use a $40 \times 40$ kpc$^2$ box, which is well matched to the field of view (FoV) of MUSE for local galaxy observations, but for the purposes of studying the evolution of the correlation function over cosmological times (\autoref{sec:res}) we instead use a smaller $20\times20$ kpc$^2$ region to avoid contamination by mergers and fly-bys, which are much more common at higher redshift; we discuss the choice of box size in more detail in \aref{app:box_size}. 

Our first step is to resample the star-forming gas cells within the box into $125\times125\times125$ pc$^3$ cells. We then convert the 3D box into a 2D map by integrating the element (hydrogen and oxygen) mass in all the cells along a line of sight normal to the galactic plane. We obtain a face-on oxygen metallicity map in a commonly used form from the definition of metallicity, $12 + \log(N_{\rm O}/N_{\rm H})$, where $N_{\rm O}$ is the column density of oxygen nuclei. The end result is a projected metallicity map with a spatial resolution of 125 pc. To complement this we also produce a star formation rate map at the same resolution by projecting the total masses of star-forming gas parcels and dividing by the star formation timescale $\tau_\mathrm{SF}$ used in the simulations (see \autoref{sec:desc}).

To quantitatively compare multiple metallicity maps, we extract two-point correlations from metallicity maps, and estimate corresponding correlation lengths. We do this in several steps, following the procedure outlined by \citet{Li21, Li23}. First, we subtract the radially-averaged metallicity in each annulus to obtain a metallicity residual map. Next we compute the two-point correlation function of the residual map, which we fit with the functional form for the two-point correlation function predicted by the \citetalias{KT18} model. This model predicts the correlation function in terms of two free parameters: injection width ($w_{\rm inj}$) and correlation length ($l_{\rm corr}$). Injection width describes the size of the initial bubble formed in explosion events (e.g., supernovae), and is usually too small \citep[$\lesssim 100$ pc;][]{Li23} to measure in either observations or cosmological simulated galaxies. Correlation length describes a characteristic length of ISM chemical mixing and is a key parameter of the \citetalias{KT18} model. We fit the simulation correlation function to the KT18 functional form using a least squares approach with $w_\mathrm{inj}$ and $l_\mathrm{corr}$ as free parameters. We refer to fits on the pure simulation data as the ``true'' values to distinguish them from the values derived from mock observations as we describe next.

\section{Mock observations in the local Universe}
\label{sec:mock}

Before using the simulations to study the evolution of metallicity correlations over cosmic time, we first confirm that the simulations can reasonably reproduce the metallicity distributions of galaxies at $z=0$ derived from observations \citet{Li21, Li23}. In this section we create mock observations of the Auriga galaxies for direct comparison with previous results, using realistic observational effects (e.g. noise, PSF, spatial sampling) drawn from the AMUSING++ survey \citep{Lopez-Coba20}. This survey uses MUSE observations, and at the mean 129 Mpc distance of AMUSING++ targets, MUSE's $1^\prime\times 1^\prime$ field of view corresponds to $40\times40$ kpc$^2$, which motivates our choice of region to extract from the the Auriga snapshots in \autoref{sec:data3}. For the purpose of this comparison, we select the six Auriga galaxies at level 3 (high spatial resolution) since they provide the best intrinsic spatial resolution, and thus the most stringent test of the effects of beam-smearing. We list the properties of these six galaxies in \autoref{tab:mock}.

We describe the pipeline we use to create the mock observations in \autoref{sec:mock_ppl}, and provide details of our noise model in \autoref{sec:noise}. We then demonstrate that the mock observations faithfully recover the real correlation lengths in the simulations (\autoref{sec:mock_vs_true}), and that these correlation lengths are in reasonable agreement with the values expected for $z=0$ galaxies (\autoref{sec:mock_vs_obs}).

\begin{table*}
\caption{Global properties and correlation lengths (in both the original simulations and mock observations for interior comparison) of the six Auriga simulations at level 3 (high spatial resolution). Columns are as follows: (1) Auriga ID; (2) stellar mass; (3) star formation rate from stellar cells in the recent 30 Myr; (4) star formation rate from star-forming gas cells, SFR$_{\rm g} = x_{\rm c} m_{\rm g} / t_{\rm sf}$, assuming a typical mass fraction of cold star formation clouds of $x_{\rm c} = 0.9$ (see details in the bottom panel of \citeauthor{Springel03}'s Fig. 1) and a star formation timescale $t_{\rm sf} = 2.2$ Gyr \citep{Grand17}; (5) half-stellar-mass radius (effective radius); (6) correlation length in the original simulation; (7) correlation length in the mock observation. For correlation lengths the central value is the 50th percentile of the posterior PDF, and the error bars show the 16th to 84th percentile range.}
\begin{tabular}{cccccccc}
\hline
Auriga ID & $M_*$ & SFR$_*$ & SFR$_{\rm g}$ & $R_e$ & $l_{\rm corr, true}$ & $l_{\rm corr, mock}$ \\
& ($10^{10}$M$\odot$) & (M$\odot$ yr$^{-1}$) & (M$\odot$ yr$^{-1}$) & (kpc) & (kpc) & (kpc) \\
(1) & (2) & (3) & (4) & (5) & (6) & (7)\\
\hline
6 & 7.0 & 2.7 & 2.2 & 5.2 & 1.873 & $1.351^{+0.016}_{-0.016}$ \\%[1.25ex]
\\
16 & 10.6 & 4.2 & 2.9 & 11.9 & 1.990 & $3.388^{+0.068}_{-0.066}$ \\%[1.25ex]
\\
21 & 10.2 & 7.1 & 6.1 & 8.8 & 2.377 & $2.642^{+0.021}_{-0.019}$ \\%[1.25ex]
\\
23 & 9.5 & 4.6 & 4.2 & 8.5 & 1.133 & $1.199^{+0.012}_{-0.011}$ \\%[1.25ex]
\\
24 & 9.7 & 4.5 & 3.2 & 9.4 & 1.554 & $1.312^{+0.022}_{-0.021}$ \\%[1.25ex]
\\
27 & 11.0 & 4.2 & 3.8 & 7.1 & 1.263 & $1.741^{+0.017}_{-0.018}$ \\%[1.25ex]
\hline
\label{tab:mock}
\end{tabular}
\end{table*}

\subsection{Pipeline for mock observations}
\label{sec:mock_ppl}

Given our extracted region, we generate mock observations using a pipeline that consists of the following five steps; the final three of these steps closely follow the analysis method described by \citet{Li23}. We also illustrate these steps in \autoref{fig:mock_obs}, using the simulated galaxy Au6 as an example.
\begin{enumerate}
    \item In order to replicate observational errors on the simulated metallicity map, we first create mock emission line maps for the lines H$\alpha$ and [N\textsc{ii}]$\lambda6584$, required by the commonly-used \citet{PP04} metallicity diagnostic (hereafter PPN2). We choose PPN2 because it is one of the simplest diagnostics and requires only two emission lines. To produce synthetic maps in the two required lines we first derive an H$\alpha$ map from the star formation rate map using the calibration suggested by \citet{Kennicutt12}, $\mbox{SFR}/(\mathrm{M}_\odot\mbox{ yr}^{-1}) = 5\times 10^{-42} L_\mathrm{H\alpha}/(\mbox{erg s}^{-1})$. Next, we use the metallicity map and the PPN2 diagnostic to predict the [N\textsc{ii}]$\lambda6584/\mbox{H}\alpha$ ratio in each pixel by solving the equation $12 + \log($O/H$) = 9.37 + 2.03x + 1.26x^2 + 0.32x^3$, where $x=$[N\textsc{ii}]$\lambda6584/\mbox{H}\alpha$ in each pixel. Multiplying the resulting value of $x$ by the H$\alpha$ map produces an [N\textsc{ii}]$\lambda 6584$ line map. We show the resulting H$\alpha$ and [N\textsc{ii}] line maps in the left column, top two rows of \autoref{fig:mock_obs}, along with the true metallicity map (left column, bottom row). Empty pixels correspond to locations where the simulations include no fluid elements dense enough to be star-forming.
    \item Next we convolve the simulated flux maps with a Gaussian PSF and add noise. We convolve the two emission line maps using a Gaussian beam with a full width at half maximum (FWHM) of $1\farcs0$, which is the median PSF size for the AMUSING++ sample \citep{Li23}. We then generate a noise map using the method described in \autoref{sec:noise}, and add this to the beam-convolved line maps. At this point our maps represent a reasonable facsimile of the data products to which we have access from the observations; we illustrate these line flux maps, together with the metallicities one would derive by directly applying the PPN2 diagnostic to them, in the second column in \autoref{fig:mock_obs}. Blank pixels in the metallicity map correspond to locations where, as a result of noise, the flux of either the H$\alpha$ or [N\textsc{ii}] line is negative and thus it is not possible to derive a metallicity.
    \item Our third step is to mimic the data quality cuts that \citet{Li23} apply to the AMUSING++ data for our simulated maps; they show that these quality cuts are required in order to extract a reliable correlation length from the metallicity field, since in their absence the correlation is corrupted by noise. Specifically, we divide our simulated line and noise maps from step (ii) to produce maps of signal-to-ratio (S/N) for both lines, and we mask pixels where the S/N is below 3. We show the masked maps in the third column of \autoref{fig:mock_obs}.
    \item Again following the procedure described in \citet{Li23} we apply the adaptive binning algorithm \textsc{adabin} to the [N\textsc{ii}]$\lambda6584$ map, which is the weaker of the two lines in the diagnostic. This algorithm recursively groups pixels into larger and larger blocks in order to increase the S/N; each region is coarsened until it reaches the target S/N of 3. We refer readers to \citet{Li23} for full details of the algorithm. Once we have re-binned the [N\textsc{ii}] map, we reconstruct the H$\alpha$ map with the same binning in order to ensure that we only ever compute line ratios at matched spatial resolution. We show the binned maps and the metallicity map derived from them in the fourth column of \autoref{fig:mock_obs}.
    \item Our final step is to mask the adaptively binned maps to avoid computing metallicities in locations that go beyond the true boundaries of ionised gas emission in the target galaxy. To achieve this, we mask pixels where in the original, unbinned H$\alpha$ map, the fluxes are detected at S/N $<3$. We show the final, masked maps in the final column of \autoref{fig:mock_obs}.
\end{enumerate}

We estimate the true correlation length in the \citetalias{KT18} model on the metallicity maps at step (i) using a least-squares fit, as discussed in \autoref{sec:data3}. For the metallicity map at step (v) we instead follow \citet{Li23} by adopting an MCMC approach that includes two additional parameters to describe observational nuisance effects: beam size $\sigma_{\rm beam}$ and the noise factor $f$. Beam size has a Gaussian prior, centred at the known beam size and with dispersion varying in different observations; its purpose in the model is to account for the fact that beam-smearing introduces artificial correlations in the inferred metallicity at small scales. The $f$ factor accounts for the decrease of two-point correlations at non-zero separations caused by noise. Its effect is to reduce the correlation at non-zero separations by a factor of $f$, which, as \cite{Li21} show, is how noise affects a measured two-point correlation function. In an extreme case where the noise is significantly larger than signal, the two-point correlation function approaches a $\delta$ function, the two-point correlation function of pure noise.

\subsection{Noise estimates}
\label{sec:noise}

The mock pipeline described in the previous section requires estimates of the noise level in the line flux maps. Here we describe the process by which we construct these maps. We first choose four galaxies from AMUSING++ \citep{Lopez-Coba20} spanning distances ranging from 127 Mpc to 131 Mpc, where the spatial resolution of MUSE ($0\farcs2$) matches the 125 pc resolution of the Auriga simulations; the chosen galaxies are SDSSJ102131.91+082419.8 (labeled as ASASSN14ba in the AMUSING++ catalogue), MCG-03-07-040 (labeled as SN2005lu), ESO584-7 (labeled as SN2007ai), and NGC539 (labeled as SN2008gg). For each of these galaxies we extract one spectrum per spaxel, from which we obtain the flux intensity and its uncertainty for both the H$\alpha$ and [N\textsc{ii}]$\lambda6584$ emission lines. In total this yields $4\times320\times320 \approx 4\times 10^5$ distinct fluxes and uncertainties, which we use to estimate the typical noise properties given their exposure time ($\sim1$ hour).

\autoref{fig:signal_noise} shows the distribution of signal and noise for these spaxels. To model the relation between signal and noise at the given distance and spatial resolution, we fit the noise as a function of the signal with a function of the form
\begin{equation}
    N = \left\{
    \begin{array}{ll}
    N_0, & S < S_0 \\
    N_0 (S/S_0)^{1/2}, & S \geq S_0
    \end{array}
    \right.,
\end{equation}
where $N_0$ and $S_0$ are parameters to be fit. The motivation for this functional form, which is consistent with the distribution shown in \autoref{fig:signal_noise}, is that a roughly constant background noise level dominates when the signal is weak, while Poisson noise should dominate when the signal is strong. Performing a simple least-squares fit of this model to the measured S/N data yields best-fit values
\begin{eqnarray}
    N_0 & = & (5.1, 4.6) \times 10^{-19}\mbox{erg s}^{-1}\mbox{ cm}^{-2}
    \\
    S_0 & = & (6.5, 1.0)\times 10^{-18}\mbox{erg s}^{-1}\mbox{ cm}^{-2},
\end{eqnarray}
where the first number in parentheses is for the H$\alpha$ line and the second for the [N\textsc{ii}] line. We show these fits by the solid lines in \autoref{fig:signal_noise}.

To generate our synthetic noise maps, we use these fits to predict the noise level in every pixel of both the H$\alpha$ and [N\textsc{ii}] maps, and then we draw a noise value for that pixel from a Gaussian distribution with zero mean and a dispersion equal to the noise level.

\begin{figure*}
\includegraphics[width=1.0\linewidth]{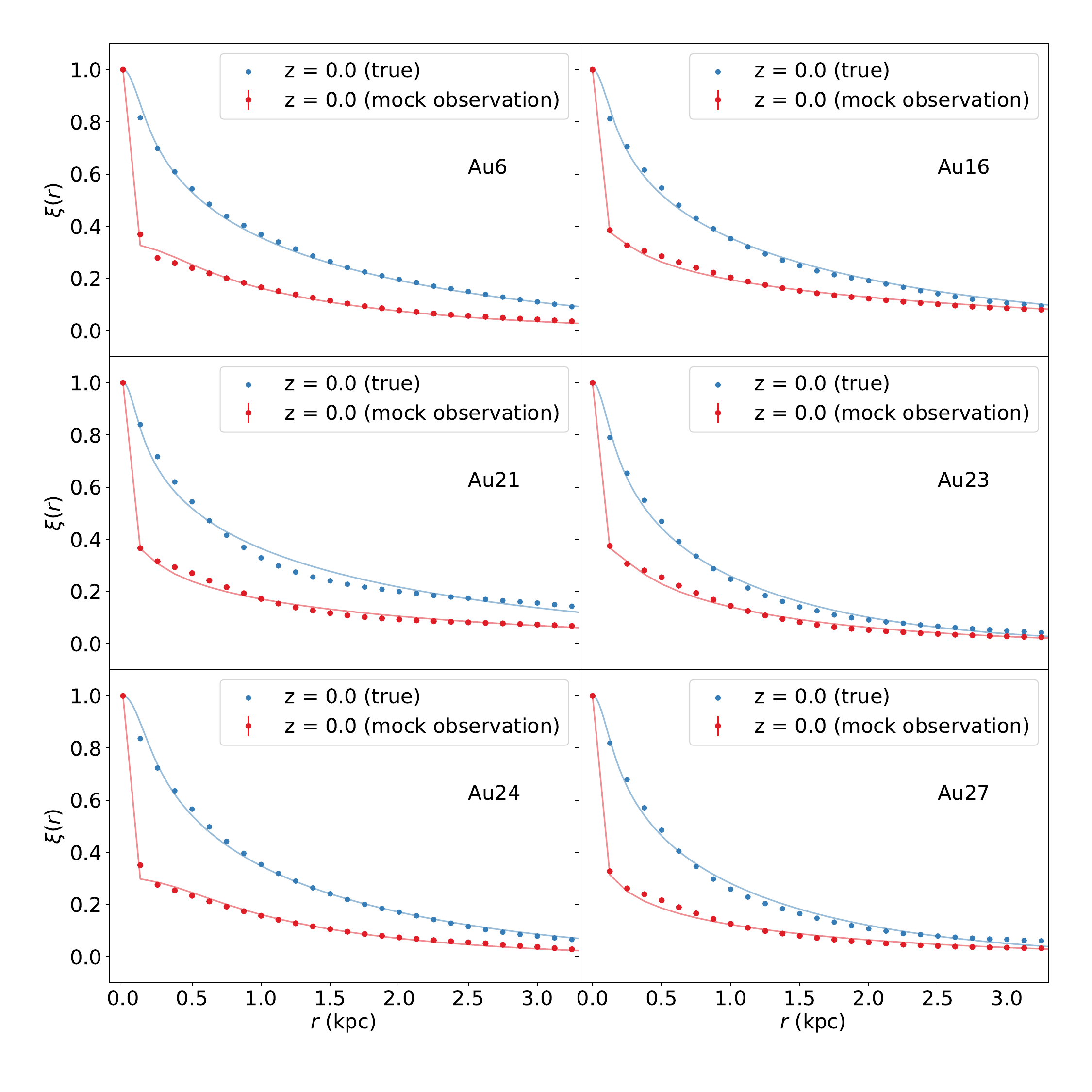}
\caption{The two-point correlation functions of the six example simulated galaxies, and the comparison between the two-point correlation functions of the pure simulations (blue dots) and those of mock observations (red dots). The blue solid lines and red solid lines represent the best estimate of the parameters in the \citetalias{KT18} model. Specifically the effects from the mock observations are shown as the discontinuity between the first two red dots.}
\label{fig:tpcf}
\end{figure*}

\begin{figure*}
\includegraphics[width=1.0\linewidth]{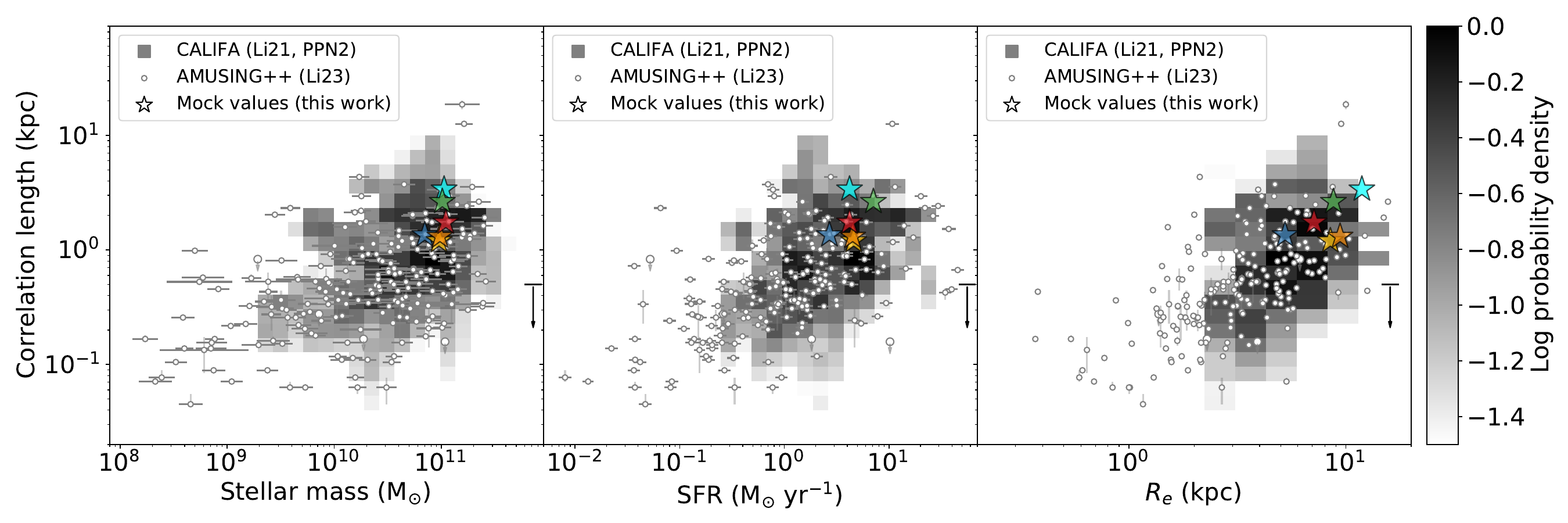}
\caption{Correlation length versus stellar mass (left), SFR (middle), and $R_e$ (right). The blue, cyan, green, yellow, orange, and red (in some panels hidden by the other two symbols) stars show the correlation lengths from the mock observations of the six halos, Au6, Au16, Au21, Au23, Au24, and Au27, respectively. The background heat map shows the distributions derived from 100 galaxies in the CALIFA survey \citep[][using the same PPN2 metallicity diagnostic]{Li21} and the circles show those of 219 galaxies in the AMUSING++ survey \citep{Li23}. The upper limit on the bottom right shows a scale below which the physics is modified by numerical smoothing at level 3 in the Auriga simulations.}
\label{fig:par}
\end{figure*}

\subsection{Comparison between mock observations and true correlation lengths}
\label{sec:mock_vs_true}

We compare the correlation lengths present in the original metallicity maps to those we recover from mock observed metallicity maps (the bottom left and the bottom right panels in \autoref{fig:mock_obs}); our goal is to establish that the correlation lengths recovered from observed metallicity maps are reasonably accurate estimates of the true correlations, despite the effects of finite telescope resolution and sensitivity. In \autoref{fig:tpcf} we show the two-point correlation functions of the both the true and mock-observed metallicity maps for six example halos. \autoref{fig:tpcf} illustrates that making a mock observation of the simulation results in suppressing the two-point correlations at all separations larger than zero due to noise, which artificially de-correlates the true metallicity map. It is a common phenomenon that appears when comparing theoretical two-point correlations and real ones. Due to inevitable noise, the two-point correlation function of an observed metallicity map will deviate from that of the true metallicity map. The overall effect of noise is to decrease the correlation function by a constant factor at all separations larger than zero. The $f$ factor in the MCMC fit captures this effect, which is why it is critical to include it.

In \autoref{tab:mock} the column $l_{\rm corr, true}$ reports the true correlation lengths measured directly from the simulations; these are reported without errors, because they come from the $\chi^2$ fitting and we do not have the uncertainties of the original metallicity maps. The column $l_{\rm corr, mock}$ reports the result from the MCMC fitting, and the reported uncertainties correspond to the 16th to 84th percentile range. \autoref{tab:mock} demonstrates that in most cases the value of $l_\mathrm{corr}$ derived from the mock observations are within $\pm 40\%$ of the true ones. We can therefore have confidence that the correlation lengths returned from mock observations are reasonably close to reality.

\subsection{Comparison between mock observations and real observations}
\label{sec:mock_vs_obs}

To examine how well our sample Auriga galaxies compare to observations, we show the relationship between $l_{\rm corr}$ (from mock observations) and $M_*$, SFR, and $R_e$ for the six example halos in \autoref{fig:par}. The results demonstrate that the Auriga simulations are consistent with observational results, in that the Auriga simulations at $z=0$ have correlation lengths well within the range observed for galaxies of similar properties. We warn that, because all the Auriga galaxies are chosen to be Milky Way analogues, this test covers only a limited dynamic range in galaxy properties. However, it is interesting to note that, despite this limited dynamic range, we do recover hints of the correlations seen in the real data, i.e., the Auriga halos with the largest stellar mass, SFR, and effective radius also tend to have larger correlation lengths.

In \autoref{fig:par} we also indicate by downward arrows our estimate of the metal injection scale in a level 3 Auriga simulation, which represents the scale below which the correlation length will be dominated by numerical resolution; we see that all the measured simulation correlation lengths lie well above this value. To estimate this scale, we first note that in the Auriga simulations metals are injected into the $4^3$ cells closest to a supernova event, so the characteristic size of a metal injection region is $4 l_{\rm g}$, where $l_{\rm g}$ is the length scale of a gas cell. Cell sizes in Auriga are adaptive, so $l_{\rm g}$ is not fixed, but we can estimate it by considering a gas cell with a density at the star formation threshold $\rho_0 = 0.13$ H / cm$^{-3}$ (mentioned in \autoref{sec:desc}) and with the typical baryonic mass resolution at level 3 of the simulations, $m_{\rm b} = 6\times10^3$ M$_{\odot}$ \citep{Grand17}. Since $\rho_0 l_{\rm g}^3 = m_{\rm b}$, a typical cell satisfying these conditions has $l_{\rm g} \sim 125$ pc, and thus we estimate the metal injection scale to be $\sim 500$ pc.

\begin{figure*}
\includegraphics[width=1.0\linewidth]{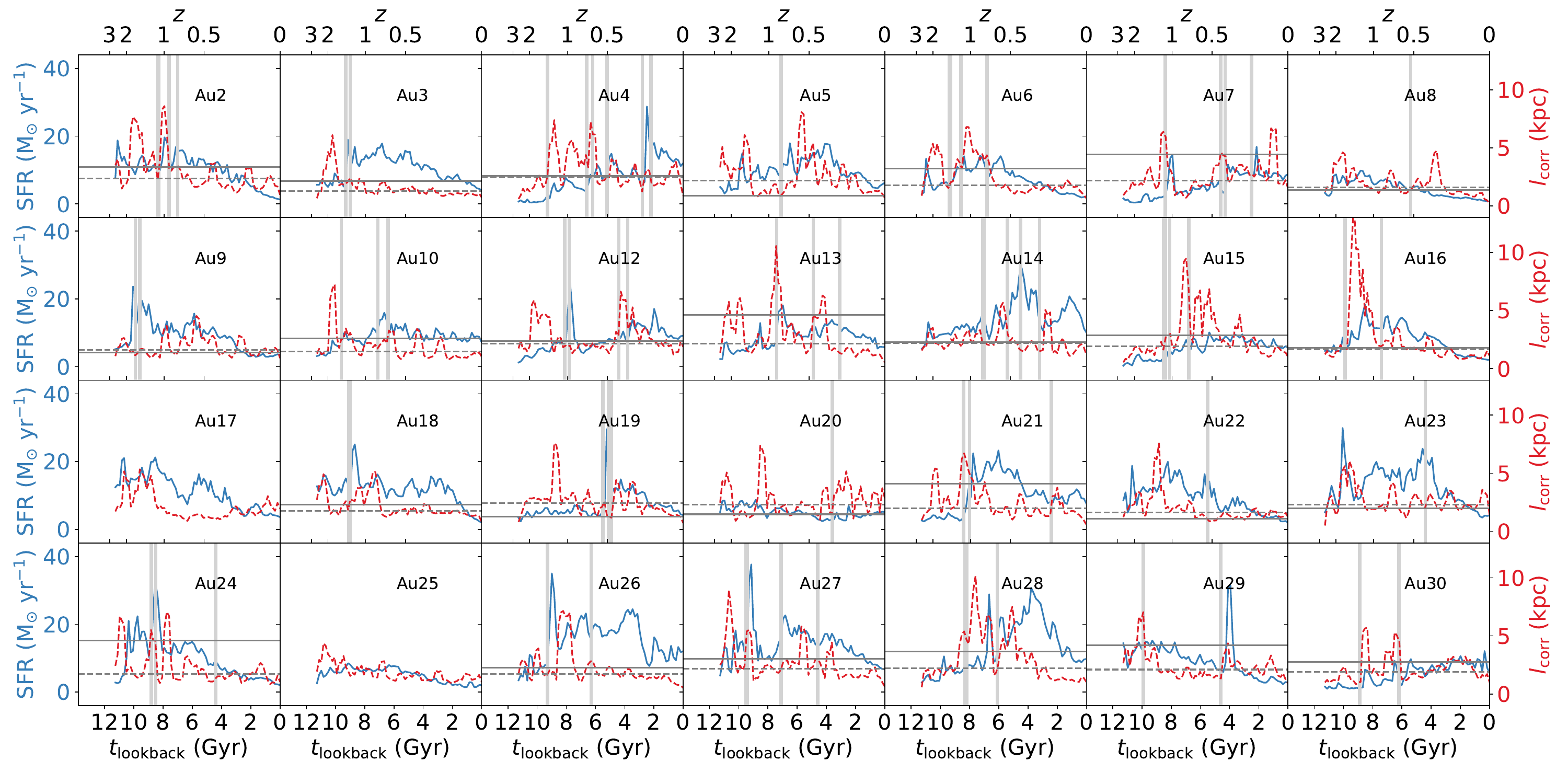}
\caption{Cosmological evolution of star formation rates (blue solid lines) and correlation lengths (red dashed lines) for all the halos. The grey vertical bands indicate periods when major mergers take place. The horizontal solid and dashed lines represent the averaged values of correlation length during merger and non-merger periods, respectively, and are plotted only for galaxies that experience at least one major merger.}
\label{fig:cosmo_evol}
\end{figure*}

\begin{figure*}
\includegraphics[width=1.0\linewidth]{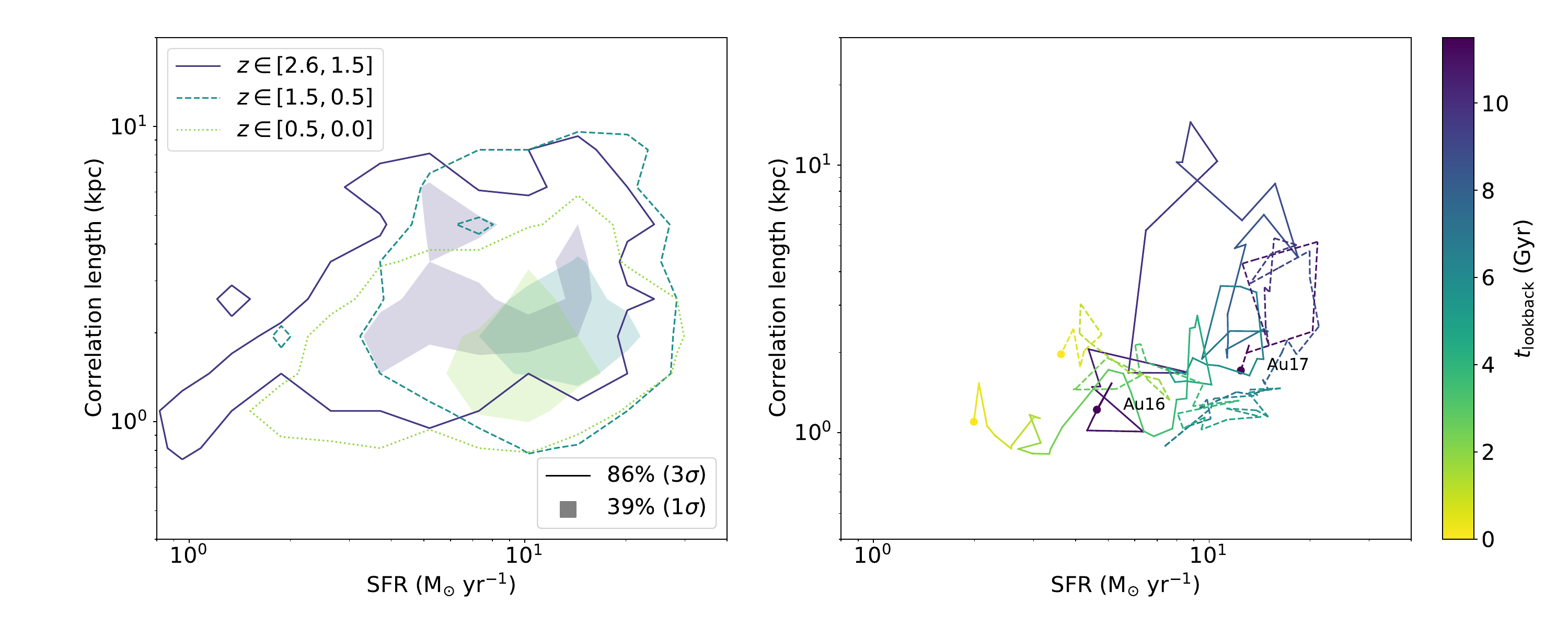}
\caption{\textit{Left}: contours showing the distribution of all the halos in the SFR-$l_\mathrm{corr}$ plane, divided up in bins of redshift. To construct this figure, we take the $l_{\rm corr}$ and SFR values from all snapshots in three temporal bins -- $z\in[2.6, 1.5]$ (cosmic noon), $z\in[1.5, 0.5]$ (disc settling), and $z\in[0.5, 0.0]$ (disc settled) -- and count the frequency in a 2D histogram in this plane with bins that are 0.14 dex wide in SFR and 0.12 dex wide in $l_{\rm corr}$. We then construct the contours shown from the 2D histograms. Each contour represents the boundary of the area within which 86\% (darker and outer) and 39\% (lighter and filled) data are included -- these contour levels correspond to $2\sigma$ and $1\sigma$ for a 2D Gaussian. The contours are coloured according to the centres of the corresponding time intervals on the colour bar to the right. \textit{Right}: the trajectory of two example halos, Au16 (solid) and Au24 (dashed), in the SFR-$l_\mathrm{corr}$ plane. Colours on the tracks indicate lookback time as shown on the colour bar to the right. The starting and ending positions are marked in circles.}
\label{fig:track}
\end{figure*}

\begin{figure*}
\includegraphics[width=1.0\linewidth]{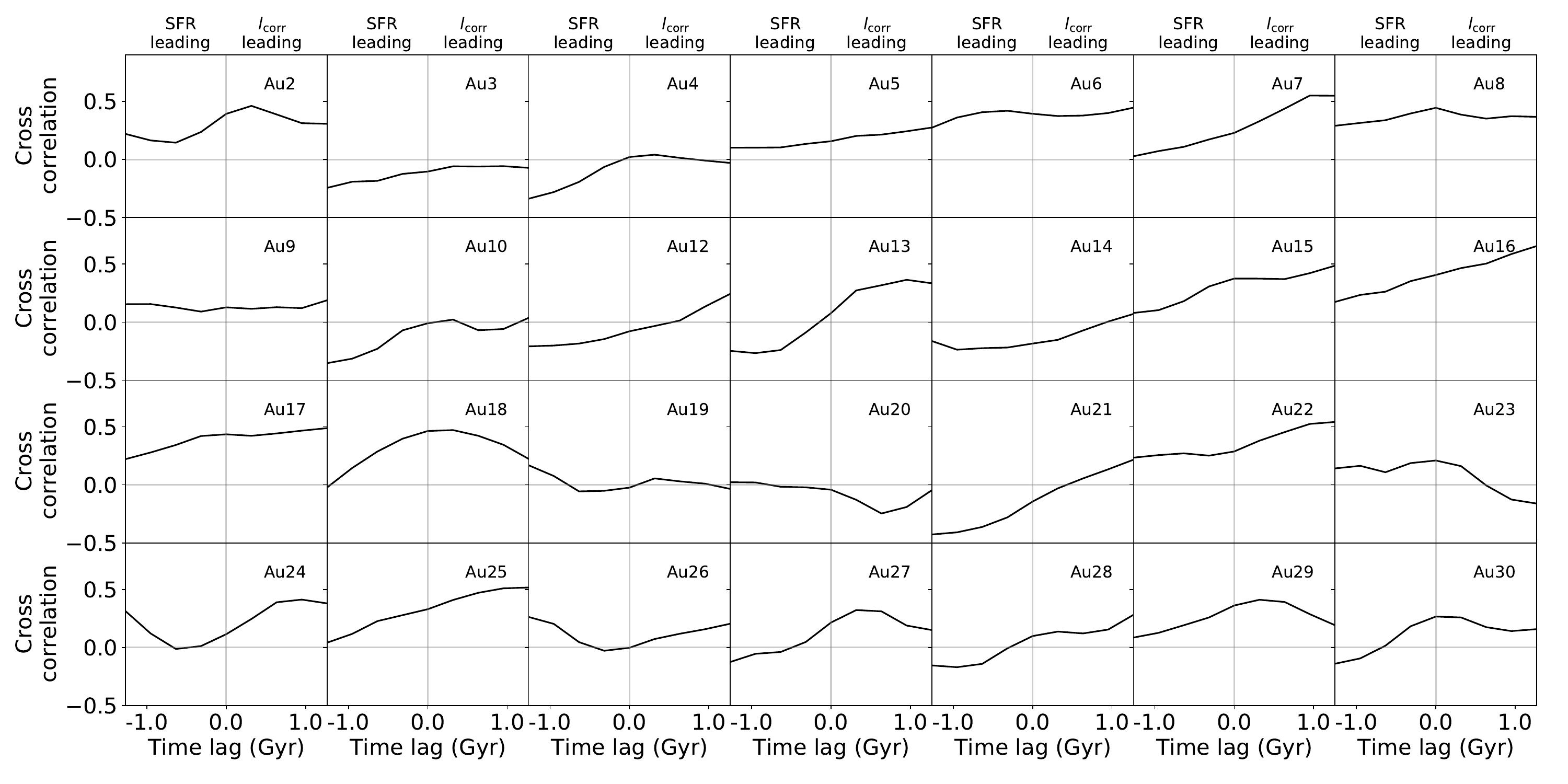}
\caption{Cross correlations between SFR and correlation lengths for all the halos. This figure is computed from \autoref{fig:cosmo_evol} and \autoref{eqn:cross_corr} (SFR as $X$ and $l_{\rm corr}$ as $Y$). Positive time lags mean that $l_{\rm corr}$ reacts ahead of SFR.}
\label{fig:cross_corr}
\end{figure*}

\section{Analysis: the evolution of metallicity correlations over cosmic time}
\label{sec:res}

Having established that the metallicity distributions in the Auriga galaxies are a reasonable match to those observed in comparable galaxies at $z=0$, we now use the Auriga sample to examine the evolution of the two-point correlation function of metallicities over cosmic time. Since we have also established that simulated observations of the Auriga simulations yield good approximations to the true correlation length, at least for observational parameters typical of current local Universe measurements, the subsequent discussion will only focus on the true two-point correlation functions, not on simulated observations -- that is, we base all analysis from this point forward on true metallicity maps such as the one shown in the first column, bottom row of \autoref{fig:mock_obs}. We have 128 such maps per Auriga halo, at time intervals of $\approx 160$ Myr. For each map of this form, we fit a noise-free \citetalias{KT18} model, following the procedure outlined in \autoref{sec:mock_ppl}, to extract a correlation length $l_\mathrm{corr}$. We therefore obtain the history of $l_\mathrm{corr}$ as a function of redshift for each Auriga galaxy.

\autoref{fig:cosmo_evol} shows the star formation histories (SFHs) and $l_{\rm corr}$ evolution over cosmic time for 28 Auriga halos at level 4 resolution. Correlation lengths are smoothed in the adjacent three snapshots (corresponding to intervals of $\approx \pm 160$ Myr on either side of a given time). We also indicate as grey vertical bands in the figure periods of time when massive mergers are taking place, using the merger list identified by \cite{Gargiulo19}. The intervals shown indicate all merger events between the primary Auriga galaxy being simulated and satellites of mass $M_{\rm sat} > 10^{10}$M$_{\odot}$\footnote{\cite{Gargiulo19} only provide lists of mergers with a fixed mass cutoff, rather than a fixed mass ratio. However, since all the Auriga halos have a final stellar mass $M_* \sim 10^{11}$ M$_{\odot}$, this choice corresponds roughly to a 10:1 mass ratio, near the commonly-adopted threshold for a minor merger.}. The width of each grey band is 160 Myr, close to the interval between two snapshots.

As shown in \autoref{fig:cosmo_evol}, correlation length fluctuations correlate roughly with star formation rate fluctuations, but in general are of larger amplitude. Nor is the correlation perfect: while fluctuations of the correlation length in some galaxies resemble amplified versions of the SFR fluctuation, for example in Au16, there are other galaxies where trends in $l_{\rm corr}$ and SFR appear to be weakly correlated at best, for example in Au3 and Au19. During merger periods correlation lengths tend to be larger on average than during non-merger periods; we illustrate these averages as solid horizontal lines for merger periods and dashed horizontal lines for non-merger periods. This is consistent with the findings of \cite{Li21} and \cite{Li23}, who note that correlation lengths tend to be larger in observed galaxies that show signs of being in the process of merging. This increase in correlation length during merger periods is particularly pronounced in some galaxies, for instance Au7, Au21, Au29, and Au30. SFR is also highly sensitive to mergers, which can trigger shock waves through tidal forces, and the common response to $l_\mathrm{corr}$ and SFR to merger events clearly accounts for at least some of the correlation between them; Au24 and Au27 provide clear examples. However, it is also clearly not always the case that mergers induce large changes in SFRs or correlation lengths, and sometimes one responds but not the other. For example both Au18 and Au19 show enhanced SFRs during mergers, but no corresponding increases in correlation length. The examples above demonstrate that neither SFR nor $l_{\rm corr}$ are perfect indicators of merger events likely due to the complex nature of merger (e.g. mass ratio, gas mass ratio, orientation).

To explore the co-variation of correlation length and SFR over cosmic time further, in the left panel of \autoref{fig:track} we show a contour plot of the distribution of galaxies in the correlation length-SFR plane broken up into in three temporal bins: high- (purple), medium- (cyan), and low-$z$ (green). The quantity shown is the fraction of lookback time that the galaxy spends at a given combination of SFR and $l_\mathrm{corr}$; thus for example the outer purple 86\% contour in the figure indicates that, if one chooses a random Auriga galaxy at a random lookback time corresponding to the redshift range $z\in [2.6, 1.5]$ and plots its coordinates $(\mathrm{SFR}, l_\mathrm{corr})$, there is an 86\% chance that these coordinates will lie within the contour. From high to low redshift, we see that the contours are largely consistent. The distribution of the Auriga galaxies in SFR-$l_\mathrm{corr}$ does not shift significantly over cosmic time, and in general resembles the sequence traced out in this plane by observed $z=0$ galaxies (c.f.~the middle panel of \autoref{fig:par}). Thus the (very broad) SFR-$l_\mathrm{corr}$ relationship appears to be essentially invariant over redshift, at least within the set of galaxies sampled by Auriga, and the evolution of individual galaxies appears to consist primarily of wandering about this relationship. At first the fact that there is no overall secular trend in where galaxies live on the SFR-$l_\mathrm{corr}$ relationship with redshift might seem surprising, but we remind the reader that what we are plotting here is the histories of particular halos that have been selected specifically to be Milky Way-analogues at $z=0$, \textit{not} the distribution of all galaxies. 

However, we caution that the Auriga simulations are by construction all Milky Way analogues, and thus cover a limited range of stellar mass and morphology. We cannot determine with certainty if this conclusion will continue to hold over a wider range of galaxy properties. However, we conjecture that it will, since the observations do include galaxies with wide range of masses and morphologies at z = 0, and these appear to be part of the same continuous distribution as Milky Way-like galaxies, with no strong dependence on morphology or other galaxy structural parameters \citep{Li21}.

The right panel of \autoref{fig:track} shows the tracks of two example halos, Au16 and Au24, in the SFR-$l_\mathrm{corr}$ plane, with time flowing from purple dots to yellow dots as indicated by the colour bar. The figure demonstrates that, while on average galaxies circulate rather than migrating systematically toward one end or the other of the SFR-$l_\mathrm{corr}$ relationship, they do circulate with a clear pattern, one that is replicated in many other Auriga halos as well, though we show only two in this figure for clarity. The pattern is that galaxies do not move in both $l_\mathrm{corr}$ and SFR simultaneously. Instead, $l_{\rm corr}$ increases first (corresponding to upwards movement in the figure) and SFR increases next (rightwards movement). The same trend is visible when both of them decrease - $l_{\rm corr}$ decreases first (downwards movement) and SFR decreases next (leftwards movement). Consequently, the track forms a roughly clock-wise cycle in the $l_{\rm corr}$ versus SFR plane.

To confirm this visual impression, we examine the cross-correlation of SFR and $l_\mathrm{corr}$, defined in the usual way whereby the cross-correlation of two time sequences $X(t)$ and $Y(t)$ is given by
\begin{equation}
\zeta(\tau) = \frac{\langle(X(t)-\langle X\rangle)(Y(t+\tau)-\langle Y\rangle )\rangle}{\sqrt{(\langle X^2\rangle - \langle X\rangle^2)(\langle Y^2\rangle - \langle Y\rangle^2)}},
\label{eqn:cross_corr}
\end{equation}
where $\langle\cdot\rangle$ denotes an average over $t$.
A positive time lag means that $Y$ reacts first and $X$ follows. \autoref{fig:cross_corr} shows the cross-correlations between SFR (as $X$) and $l_{\rm corr}$ (as $Y$) for all the halos. In most cases, we see that SFR and $l_{\rm corr}$ are positively correlated at zero time lag, consistent with the overall positively-sloped ``main sequence'' visible in the SFR versus $l_{\rm corr}$ plane. However, it is also clear that in most cases the cross-correlations have a positive slope, meaning that the correlation is stronger at positive time lag, corresponding to $l_{\rm corr}$ changing first and the SFR reacting slightly later.

This is an interesting phenomenon because we see for the first time the causality of the interplay between $l_{\rm corr}$ and SFR. This cannot solely be merger-driven, since we see the same general positive trend in the cross-correlation function for the two halos that have no mergers (Au17 and Au25) as for all the others. The positive time lag is probably a result of the correlation length and SFR reacting to perturbations on different time scales -- the natural response time for $l_{\rm corr}$ is a galactic orbital period, while the natural response time for the SFR is the SFR timescale of $\sim 2$ Gyr, which is much longer than an orbital period. This explains both why $l_{\rm corr}$ fluctuates more, and why its fluctuations tend to lead SFR fluctuations.

We note that, while the SFR timescale of $\sim 2$ Gyr is hardwired into the Auriga star formation prescription, and the $\sim 200$ Myr orbital timescale is implicitly fixed by the choice to simulate Milky Way analogues, the general physical point we make here is true more generally: all observed galaxies have star formation timescales roughly an order of magnitude longer than their orbital times. This statement is equivalent to the second form of the well-known \citeauthor{Kennicutt98} relation ($\Sigma_\mathrm{SFR}$ vs. $\Sigma_\mathrm{g}/t_\mathrm{dyn}$, where $\Sigma_\mathrm{SFR}$, $\Sigma_\mathrm{g}$, and $t_\mathrm{dyn}$ denote star formation rate surface density, gas surface density, and local dynamical timescale, respectively). The relation also holds for high-redshift galaxies \citep[e.g.][]{Daddi10}.

\section{Conclusions}
\label{sec:dis_sum}

In this study, we utilise the Auriga simulations to investigate the evolution of the spatial distributions of chemical elements in galaxies using the two-point correlation function as a tool. To quantitatively compare the two-point correlations of different galaxies, and of individual galaxies as they evolve over time, we study the correlation length ($l_{\rm corr}$) introduced in the \citet{KT18} model for 2D galactic abundance distributions. As a parameter that reveals galaxy chemical mixing mechanisms, the correlation length provides a unique window into galaxy past evolution history.

We first confirm that the Auriga simulations have correlation lengths in galactic metal fields comparable to local observation. We first carry out simulated observations of the $z = 0$ Auriga snapshots, demonstrating that when we add realistic noise and beam smearing effects, the Auriga simulations produce galaxies with correlation lengths $l_{\rm corr}$ consistent with observed values for galaxies of similar properties, indicating that the Auriga simulations capture the dominant chemical mixing processes involved in galaxy evolution. We further show that correlation lengths recovered from these simulated observations are reasonably close to the true values present in the simulations.

We find that for an individual galaxy there is no significant secular evolution in its correlation length, or in its correlation length versus star formation rate, over cosmological timescales. The mean correlation lengths of a population of galaxies at high-$z$ shows only marginal differences from the mean of those same galaxies at low-$z$, and the galaxy-to-galaxy scatter of $l_{\rm corr}$ also remains unchanged over cosmic time. We therefore conclude that the relationship between correlation length and star formation rate is \textit{not} the result of a build-up over cosmic history. Instead, it appears to be an equilibrium relationship that is established in a time much less than a Hubble time. Galaxies can move along this relationship in response to perturbations in their environments, but the relationship itself is essentially invariant (xxx). This picture is very similar to the one proposed for the origin of galaxy metallicity gradients and their relationship with other galaxy properties such as mass and star formation rate \cite{Sharda21}.

Furthermore, our analysis reveals an intriguing trend that fluctuations in correlation lengths precede fluctuations in star formation rate. This finding suggests that the scatter of the $l_{\rm corr}$ distribution might be a result of galaxies being at different evolutionary stages, whereby galactic correlation lengths react more rapidly to external perturbations than star formation rates. The former evolve on timescales comparable to the galactic orbital period, while the latter are change only over many orbital periods, so at a given time where a galaxy lies in the SFR-$l_\mathrm{corr}$ depends in part on whether its star formation rate has had time to ``catch up'' to the changes in correlation length induced by whatever perturbed it most recently.

Finally, in the Auriga simulations we have presented, although gas cells in the star-forming disc can be as small as a few tens of pc, metals are injected across 64 cells and therefore the injection width, the other key parameter of the \citetalias{KT18} model with typical values of $\lesssim 100$ pc \citep{Li23} is not resolved everywhere. However, in principle the value of $w_\mathrm{inj}$ should vary between elements with different nucleosynthetic origins (e.g., N versus O). One should be able to find signatures of this effect in both observations and higher resolution simulations. In future work we intend to apply two-point correlations to higher resolution zoom-in simulations, to investigate if they are able to recover the imprints of elements' differing nucleosynthetic origins on their present-day spatial distributions.

\section*{Acknowledgements}

RG acknowledges support from an STFC Ernest Rutherford Fellowship (ST/W003643/1). EW \& JTM acknowledge support by the Australian Research Council Centre of Excellence for All Sky Astrophysics in 3 Dimensions (ASTRO 3D), through project number CE170100013. MRK acknowledges support from the Australian Research Council through award FL220100020.

\section*{Data Availability}

The scripts and plots for this article will be shared on reasonable request to the corresponding author. The \textsc{arepo} code is publicly available in \cite{Weinberger20}.



\bibliographystyle{mnras}
\bibliography{ref} % if your bibtex file is called example.bib

%\begin{thebibliography}{99}
%\end{thebibliography}

%%%%%%%%%%%%%%%%%%%%%%%%%%%%%%%%%%%%%%%%%%%%%%%%%%

%%%%%%%%%%%%%%%%% APPENDICES %%%%%%%%%%%%%%%%%%%%%

% \appendix

\appendix

\section{Comparison of results derived from level 3 and level 4}
\label{app:levels}

\begin{figure}
\includegraphics[width=1.0\linewidth]{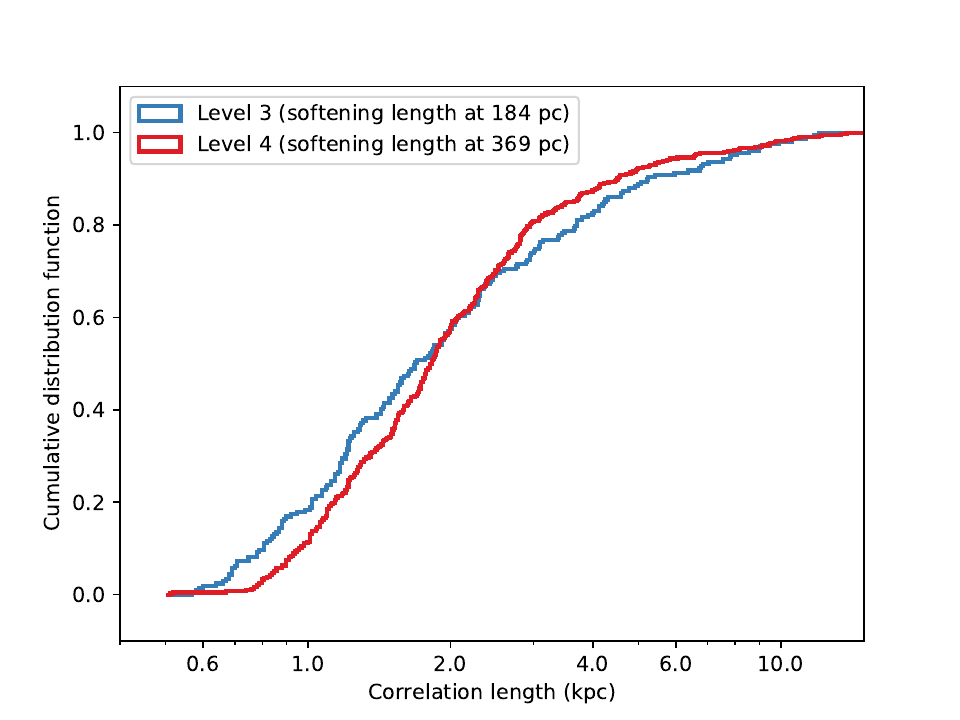}
\caption{Cumulative distribution function of correlation lengths for the six Auriga halos that are simulated at both level 3 (blue) and level 4 (red) resolution.}
\label{fig:levels}
\end{figure}

The majority of the Auriga simulations use ``level 4'' resolution, for which the maximum softening length is 369 physical pc, but a subset of six halos (Au6, Au16, Au21, Au23, Au24, and Au27) were also simulated at a resolution with a maximum softening length of 184 physical pc (``level 3'' resolution). In order to test the resolution dependence of our results, we compute correlation lengths from both sets of simulations. In \autoref{fig:levels} we show the cumulative distribution function (CDF) of correlation lengths of the six halos that are simulated at both level 3 and level 4; the CDF that we show here is the distribution of correlation lengths over all all snapshots for all six halos, and thus represents the distribution in time. The two distributions are clearly very similar, and a quantitative comparison using a two-sided KS test indicates that we cannot reject the null hypothesis that the level 3 and level 4 simulation results were drawn from the same underlying distribution ($p=0.083$). This suggests that our analysis is robust against changing resolutions.

\section{Comparison of results derived from different box sizes}
\label{app:box_size}

\begin{figure}
\includegraphics[width=1.0\linewidth]{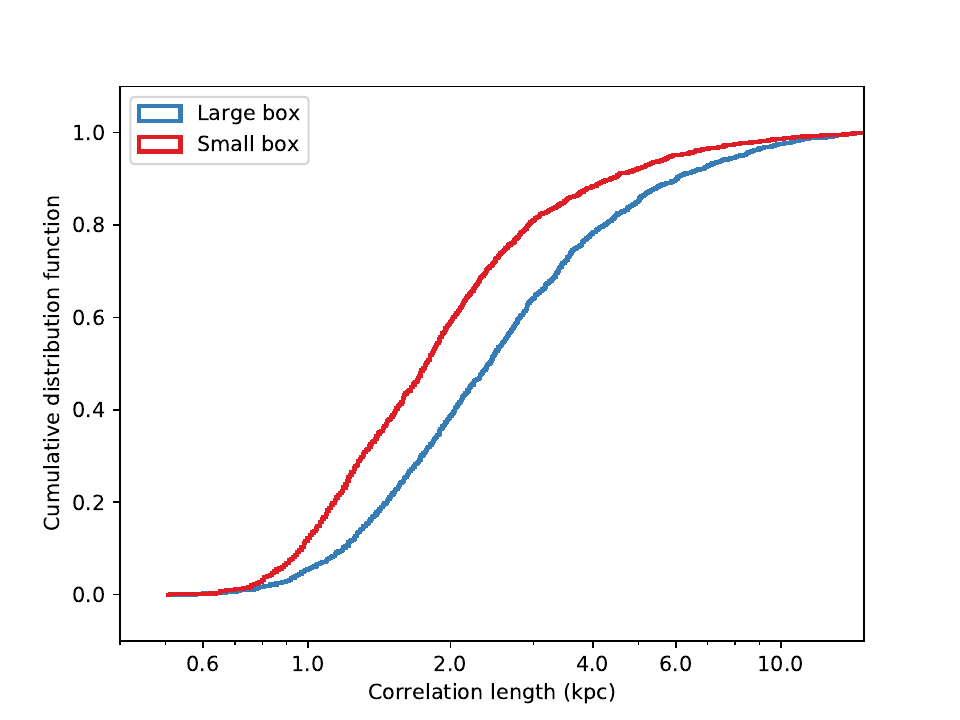}
\caption{Cumulative distribution function of correlation lengths for the 28 Auriga halos using a larger ($40\times 40$ kpc$^2$, blue) and smaller ($20\times 20$ kpc$^2$, red) FoV.}
\label{fig:box_size}
\end{figure}

\begin{figure*}
\includegraphics[width=.8\linewidth]{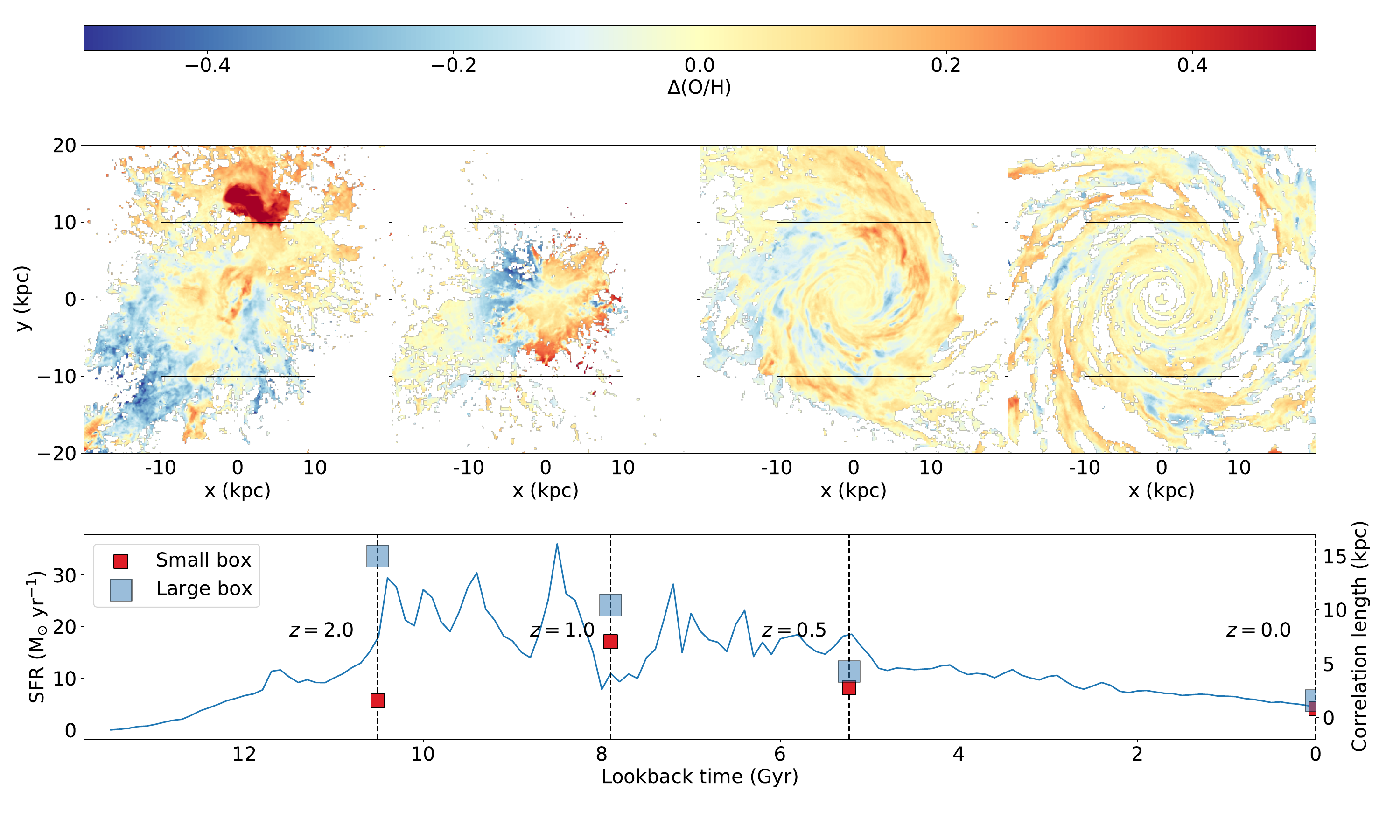}
\caption{An illustration of an example halo Au24. The first row shows the metallicity residual maps of a $40\times40$ kpc$^2$ FoV at four cosmic times ($z=2.0, 1.0, 0.5$, and 0.0 from left to right), while the black boxes indicate a $20\times 20$ kpc$^2$ FoV. The second row illustrates the star formation history in the blue curve and the correlation lengths derived from larger (blue squares) and smaller (red squares) at different cosmic times indicated using the black vertical dashed lines.}
\label{fig:box_size_Au24}
\end{figure*}

As discussed in the main text, there is some subtlety in the choice of field of view (FoV) around each galaxy to examine when computing the two-point correlation function. A larger FoV of $40\times40$ kpc$^2$ at the median $\sim129$ Mpc distance of the AMUSING++ sample is well-matched to the MUSE FoV, and thus is well-suited to making mock observations, but we find that a smaller $20\times 20$ kpc$^2$ FoV is preferable for studying cosmological evolution. To motivate that choice and examine its impact, in \autoref{fig:box_size} we show the CDF of correlation lengths of all the 28 halos at level 4 computed using both the larger and smaller FoV. We find that the larger FoV choice leads to a distribution that is shifted significantly to larger correlation lengths. The two-sided KS test indicates that we reject the null hypothesis that the larger and smaller FoV samples were drawn from different parent distributions ($p=6.6\times10^{-16}$). This suggests that the choice of box size will affect our analysis.

Given that the choice of FoV matters, it is important to understand what causes the difference. For this purpose we show a detailed example of Au24 in \autoref{fig:box_size_Au24}. In this figure we show in the upper panels the metallicity residual maps at four redshifts ($z=2, 1, 0.5$, and 0), and in the lower panel the star formation history and the correlation length as a function of redshift. We see that at the two lower redshifts, $z=0$ and $0.5$, the correlation lengths for the two different FoV sizes are nearly identical, and they differ by only tens of percent even at $z=1$. By contrast, there is a very large difference in correlation length in the $z=2$ snapshot, and examination of the metallicity maps at star formation history makes it clear why: at $z=2$ the galaxy is on the cusp of a merger, leading to a substantial increase in star formation rate. Because the merger has not yet occurred, however, we can clearly see in the larger FoV two distinct galaxies with different mean metallicities. When we then compute the two-point correlation of resulting map, we get a very large value that in effect corresponds to the projected separation between the two pre-merger galaxies, rather than describing the metal field within either of them; a similar phenomenon is seen in observations of interacting galaxies \citep{Li23}. The smaller FoV avoids this effect because it is not large enough to include gas in satellite galaxies or metal-poor gas from the circumgalactic medium (CGM). This phenomenon drives the differences in the distributions of correlation lengths visible in \autoref{fig:box_size}, and it motivates us to choose the smaller box size for our cosmological analysis because we wish to focus on correlation lengths within galaxies, and therefore to the extent possible to avoid contamination by pre-merger interacting systems.

%%%%%%%FoV and the s%%%%%%%%%%%%%%the larger %%%%%%%%%%%%%%%%%%%%%%%%%%%%%

% Don't change these lines
\bsp	% typesetting comment
\label{lastpage}
\end{document}